\documentclass{pasj00}
\draft

\SetRunningHead{Astronomical Society of Japan}{Usage of \texttt{pasj00.cls}}

\title{Deep CO Observations and the CO-to-$\mathrm{H_2}$ Conversion
Factor in DDO\ 154, a Low Metallicity Dwarf Irregular Galaxy
}
\author{Shinya \textsc{Komugi}$^{1,2}$,
Chikako \textsc{Yasui}$^2$,
Naoto \textsc{Kobayashi}$^3$,
Bunyo \textsc{Hatsukade}$^2$
Kotaro \textsc{Kohno}$^3$,
Yoshiaki \textsc{Sofue}$^4$
\and Shiori \textsc{Kyu}$^3$}

\affil{
$^1$Institute of Space and Astronautical Science, Japan Aerospace
Exploration Agency, \\
3--1--1 Yoshinodai, Sagamihara, Kanagawa, 229--8510}

\affil{
$^2$National Astronomical Observatory of Japan, 2--21--1 Osawa,
Mitaka-shi, Tokyo, 181--8588}

\affil{
$^3$Institute of Astronomy, The University of Tokyo, 2--21--1 Osawa,
Mitaka-shi, Tokyo, 181--8588}

\affil{
$^4$Dep. Physics, Meisei University, Hino, Tokyo, 191-8506}

\email{s.komugi@nao.ac.jp}


\Received{$\langle$reception date$\rangle$}
\Accepted{$\langle$acception date$\rangle$}
\Published{$\langle$publication date$\rangle$}
\begin{document}
\maketitle

\begin{abstract}
We present a deep spectroscopic search for CO emission in the dwarf
 irregular galaxy DDO154, which has an Oxygen abundance of only 1/20 the
 solar value.  The observations were conducted in order to constrain the
 CO-to-$\mathrm{H_2}$ conversion factor at low
 metallicity.  No CO was detected, however, despite being one of the sensitive
 observations done towards galaxies of this type.  We succeed in putting
 a strong lower limit on the conversion factor, at least 10 times the Galactic
 value.  Our result supports previous
 studies which argue for a high conversion factor at low metallicity.
\end{abstract}

\section{Introduction}

Tracing the molecular gas content is an essential part of
studying star formation in galaxies.  Molecular hydrogen, however,
cannot be observed directly except in hot and/or shocked environments
because of its lack of dipole moments.
Consequently, CO lines have been used via the ``CO-to-$\mathrm{H_2}$
conversion factor ($X_\mathrm{CO}$)'' to calibrate the molecular gas
quantity.  Measurements of $X_\mathrm{CO}$ are conducted mainly by
comparing the dynamical mass and the CO luminosity of a molecular cloud,
and is known to be constant (but with considerable scatter) in the Milky
way at around $X_\mathrm{CO}=3.0\times 10^{20}\ (\mathrm{cm^{-2}
(K\ km\ s^{-1})^{-1}})$ 
\citep{scoville87, strong88}.  The value of $X_\mathrm{CO}$ in other environments is
controversial, however.  In particular, the study of $X_\mathrm{CO}$ and
its dependence on metallicity in low metallicity environments have met
severe difficulties.  While
\citet{wilson95} and \citet{arimoto96} find an increasing $X_\mathrm{CO}$
with decreasing metallicity, \citet{rosolowsky03}, \citet{bolatto04},
\citet{bolatto08} and others find a constant value of $X_\mathrm{CO}$
within a range of metallicities.  It is important to note, however, that
previous studies which find a constant conversion factor are based on CO
observations at relatively high metallicities, i.e., higher than 
$12+\log[\mathrm{O/H}] \sim 8.0$.
In order to gain decisive information on the metallicity dependence of
$X_\mathrm{CO}$, observations at low metallicities is necessary.  This
has been a difficult task, however, because CO has
failed to be detected in galaxies with metallicity lower than
$12+\log[\mathrm{O/H}] \sim 8.0$ \citep{taylor01, leroy05}.  The lowest
metallicity galaxy with detected CO is IZw\ 36, at $12+\log[\mathrm{O/H}] \sim 7.9$ 
\cite{young95} which is only tentative. 
  Attempts to detect CO in low metallicity systems \citep{ohta88, wilson92, ohta93, 
verter95, taylor01,
buyle06} may have failed for two reasons; 
1) CO fails to be a good tracer of molecular gas at such metallicities (i.e., a high
conversion factor), or
2) Molecular gas is actually less abundant in such galaxies.
While the first explanation is supported from theoretical views
\citep{maloney88, sakamoto96, bell06}, the second explanation is less convincing, 
since star formation is still found in low metallicity dwarfs, and star formation 
requires gas at high densities.

In either case, calibrating $X_\mathrm{CO}$ requires an actual detection
of CO in these low metallicity systems.  We have conducted a deep search
for $^{12}\mathrm{CO}(J\ =\ 1-0)$ in the low metallicity dwarf irregular 
galaxy DDO\ 154, a member of the Local Group.

Section 2 explains the target galaxy and the observed region.  
Section 3 presents observations at
 the Nobeyama Radio Observatory (NRO) 45m telescope and the Institut
 Radio Astronomie Millimetrique (IRAM) 30m telescope, and the resulting spectra.
Section 4 discusses the $X_\mathrm{CO}$, and its upper limits.

\section{Observing Target}
\subsection{DDO\ 154}
DDO154 is a low surface brightness (LSB) dwarf irregular (dI) galaxy located at a
distance of 3.2 Mpc (\cite{carignan98}:CP98), with an optical radius
of $R_{Ho}=1.4\mathrm{kpc}$.  The metallicity of $12+\log[{O/H}] =7.67$(\cite{vanzee97},
\cite{kennicutt01}; KS01), which is about 1/20 of our Galaxy, places
DDO154 as one of the most metal-deficient galaxies.  Basic properties of 
DDO\ 154 are presented in table \ref{ddo154prop}.
DDO154 is unique in that dark matter constitutes 90\% of its dynamical
mass, therefore giving it the name ``dark galaxy'' \citep{kennicutt01}.  Stars account for
only 2\% of the dynamical mass, with $M_\mathrm{HI}/L_\mathrm{B}=8$, and 
$M_\mathrm{HI}=2.5\times 10^8 M_\odot$ \citep{carignan89}, placing
DDO154 one of the most gas rich galaxies known.  Spectroscopy by
KS01 reveals that its oxygen yield is consistent with a closed box
model, making it a candidate for one of the most unevolved Population I
systems. 
Star formation is evident from H$\alpha$ observations.  Sixteen HII regions
were identified in KS01, and the ratio of the present star formation
rate to average past star formation rate (birthrate parameter $b$) was
found to be $1/2$ to $1/3$, showing a quiescent phase.  The gas
consumption time is found to be over 200 Gyr \citep{kennicutt01}.
  
The richness of neutral gas compared to any other
previously observed low metallicity sources, and the presence of
substantial star formation requiring an ample amount of molecular gas, 
make DDO154 the optimum target for detecting
CO in a low metallicity environment.

\begin{table}[t]
  \caption{Properties of DDO\ 154 and Region 2}\label{ddo154prop} 
  \begin{center}
     \begin{tabular}{ccc} \hline \hline
     DDO\ 154       &  &  Ref.  \\ \hline    
     R.A.(J2000)     & 12 54 05.2 &  (1)  \\
     Dec.(J2000)     & 27 08 59   &  (1)  \\
     Morphology      & IB(s)m     &  (1)  \\
     Distance        & 3.2 Mpc    &  (2)  \\
     Optical Radius  & 1.4 kpc    &  (2)  \\
     $V_\mathrm{rad}$& 374 $\mathrm{km\ s^{-1}}$  & (1)  \\
     $12+\log(O/H)$   & 7.67       &  (3)  \\
     $\mathrm{M_{HI}}$& $2.5\times 10^8 \ \mathrm{M_{\odot}}$ & (2) \\
     $L_B$           & $3.1\times 10^7 \ \mathrm{L_{\odot}}$ & (2) \\
      \hline
      Region 2 & &  \\ \hline    
     R.A.(J2000)  & 12 54 03.9  &  (4) \\
     Dec.(J2000)  & 27 09 05    &  (4) \\ 
     L(H$\alpha$) & $7.0\times 10^{36}\ \mathrm{(erg\ s^{-1})}$ & (4) \\
     Radius       & $3.5^{\prime \prime}$ & (4) \\ \hline
  \end{tabular}
  \end{center}
References; (1) From the NASA Extragalactic Database (NED). (2) \citet{carignan98}.
            (3) \citet{vanzee97}. (4) \citet{kennicutt01}.
   \end{table}

\subsection{Region 2}

Of the HII regions identified in KS01, we selected
Region 2 as our target for the CO emission search. 
Figure 1 shows the 24$\mu m$, H$\alpha$ (KS01) and HST F606W band image.
Various properties of Region 2 are presented in table \ref{ddo154prop}.

Region 2 is a relatively compact HII region with a diameter of
     $7^{\prime \prime}$, corresponding to 110 pc (KS01).  The H$\alpha$
     luminosity of $L\mathrm{(H\alpha)}=7.0\times 10^{36}$ is the third most
     luminous as a single discrete HII region, and the brightest of
     those with deep spectroscopy by KS01.  This luminosity corresponds
     to a star formation rate of $1.0\times 10^{-3} \ 
M_\odot \ \mathrm{kpc^{-2} \ yr^{-1}}$, calculated from 
the formulation of \citet{kennicutt98}.  Assuming the same star
     formation law as low surface brightness galaxies in the same star
     formation rate range \citep{bigiel08}, this star formation rate
     corresponds to a molecular gas surface density of about 
$\mathrm{N(H_2)}\sim 2.5 M_\odot \mathrm{pc^{-2}}=1.6\times 10^{20} \
     \mathrm{cm^{-2}}$.  Absorption of H$\alpha$ photons by dust can be
     neglected here, based on the following estimate.  From the SINGs
     survey \citep{kennicutt03} data release 5, the 24$\mu \mathrm{m}$
     luminosity of Region 2 is 
$\nu L(24) = 7.2 \times 10^{22} \ \ \mathrm{(erg \ s^{-1})}$.  Assuming
     that the scaling between H$\alpha$ and 24$\mu \mathrm{m}$-derived
     star formation rates can be approximated as in \citet{calzetti07},
     the 24$\mu \mathrm{m}$ contribution to the true star formation rate
     is only $\sim 3 \%$.  This is in accordance with the low
     logarithmic $\mathrm{H \beta}$ extinction derived by
     \citet{kennicutt01} of $C(\mathrm{H \beta}) = 0.0$.

The negligible dust extinction may seem inconsistent with the
24$\mu \mathrm{m}$ emission and the assumed presence of
molecular gas, but the ISM generally has a clumpy
geometry as observed routinely in the Milky Way and other nearby
galaxies \citep{bally87}.  It is probable that dust has a clumpy
distribution around Region 2, so that most sightlines toward the HII
region are extinction-free, but that most of the molecular gas is
confined in a small fraction of the area hidden by
high density dust.

HST observation (see figure 1 inset) reveals an apparently associated stellar cluster
in this region.
Further, Region 2 is the only source in this galaxy that is detected both at 
24$\mu \mathrm{m}$
and H$\alpha$, suggesting that this source is a candidate for the youngest
cluster in DDO 154, with ample molecular gas.

\begin{figure}[t]
  \begin{center}
    \FigureFile(8cm,8cm,clip){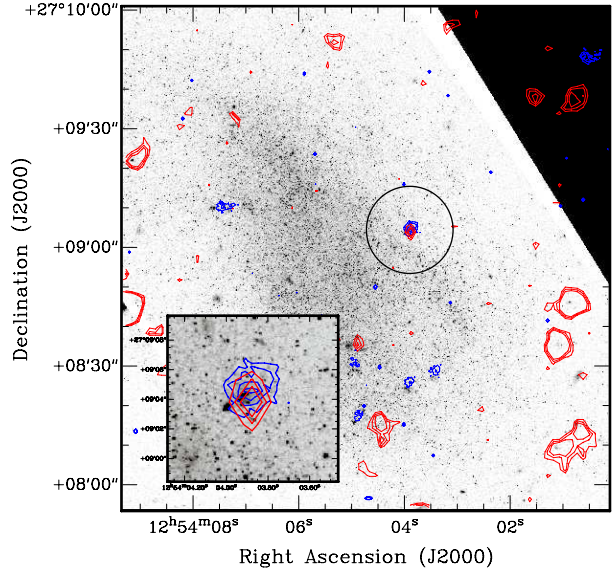}
  \end{center}
\caption{H$\alpha$ (Blue) and Spitzer MIPS 24$\mu \mathrm{m}$ (Red)
 contours overlaid on HST F606W image.  Region 2 is marked by the black
 circle, representing the IRAM 30m beamsize of $22^{\prime \prime}$. The inset
is a close-up of region 2.}
\end{figure}

\section{Observation}

\subsection{NRO 45m}
The $^{12}\mathrm{CO}(J\ =\ 1-0)$ line (rest frequency 115.271204 GHz)
 was observed with the Nobeyama Radio Observatory (NRO) 45m telescope in April, 2007.
  Two SIS receivers, S100 and/or S80, were used for the observations. 
 The backend was the 2048 channel acousto-optical spectrometer (AOS)
 with a frequency resolution of 250MHz, centered at the systemic velocity
 of DDO\ 154, $374\ \mathrm{km\ s^{-1}}$.  The CO
line was observed in the upper sideband.
Calibration was done with the standard chopper-whell method.
  Typical system temperatures ranged from 700 to 1000 K, under
 relatively bad weather conditions.  Approximately 20 hours were spent
 on source.

The beamwidth was 15$^{\prime \prime}$ at 115.3 GHz.  Pointing accuracy was checked using SiO masers or
 a continuum source (3C273) at 43GHz every 1.5-2 hours, and found to be accurate
 within 8$^{\prime \prime}$ during most of the run.

The obtained data  were reduced with the NEWSTAR software used commonly at NRO, which
 is based on the AIPS package. 
 After flagging of bad spectra, a first order baseline was subtracted and then averaged.
  The antenna temperature
 $\mathrm{T_A}^*$ was converted to main beam temperature $\mathrm{T_{mb}}$ through 
 $\mathrm{T_A^*}=\eta_{\mathrm{eff}}\mathrm{T_{mb}}$ where $\eta_{\mathrm{eff}}$ is the
 main beam efficiency,
 taken to be 0.34 based on December 2006 observations of the Saturn and 3C279.   

\begin{figure}[t]

  \begin{center}
    \FigureFile(7.5cm,9cm){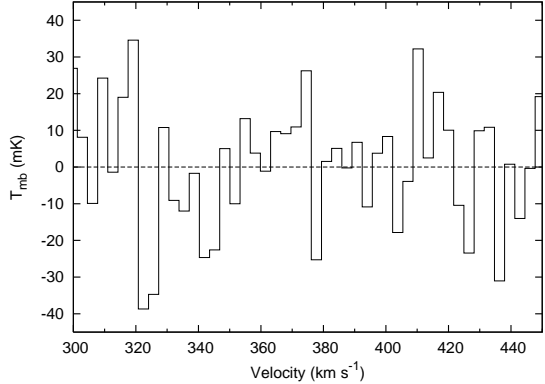}
  \end{center}
  \caption{Spectrum of Region 2 from the NRO 45m telescope.  Only the
 velocity range in the vicinity of Region 2 is shown.}\label{NROspectra}
\end{figure}

After binning the  resulting spectrum to 3.3 km/s resolution, $1 \sigma$
 r.m.s noise was 17 mK ($\mathrm{T_{mb}}$; $5.8\mathrm{mK}$ in $\mathrm{T_a^*}$).  
The spectrum is shown in figure
 \ref{NROspectra}.

No signal was detected, and we therefore conducted further
follow-up observations at the Institut de Radio Astronomie Millimetrique (IRAM) 30m 
telescope.

\begin{figure}[ht]
  \begin{center}
    \FigureFile(7.5cm,9cm){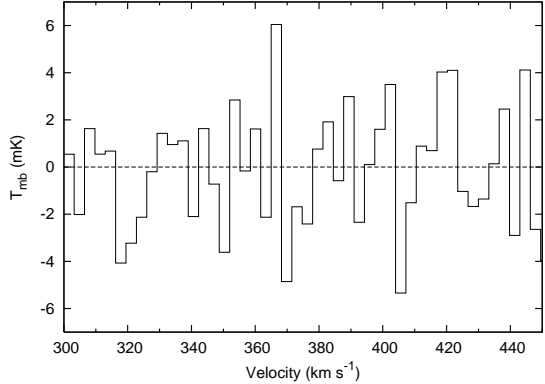}
  \end{center}
  \caption{Spectrum of Region 2 from the IRAM 30m telescope.  Only the
 velocity range in the vicinity of Region 2 is shown.}\label{IRAMspectra}
\end{figure}

\subsection{IRAM 30m}
The $^{12}\mathrm{CO}(J\ =\ 1-0)$ line (rest frequency 115.271204GHz)
 was observed with the IRAM 30m telescope in November, 2007.
Two SIS receivers, A100 and B100, were used in combination with the
 VESPA spectrometer.  The typical system temperature was 320 K.
Approximately 5 hours were spent on source.  The
 beamwidth at 115 GHz is $22^{\prime \prime}$.

The obtained data  were reduced with the CLASS software. 
 After flagging of bad spectra, a second order baseline was subtracted
 and then averaged.  The main beam efficiency $\eta_{\mathrm{eff}}$ was
 taken to be 0.7, based on March 2005 observation of Mars and Uranus.
 The spectrum was binned to give a final velocity resolution of 3.3
 km/s.  The $1 \sigma$ r.m.s. noise was 2.6 mK ($\mathrm{T_{mb}}$).  The
 spectrum is shown in figure \ref{IRAMspectra}.

No signals were detected above the 3 $\sigma$ threshold.  
The 3$\sigma$ upper limit for the CO intensity can be calculated by 
\begin{equation}
\mathrm{I_{CO}} <3 \sigma_\mathrm{rms} \sqrt{\delta V \Delta V_\mathrm{CO}}
 \quad [\mathrm{K\ km\ s^{-1}}]
\label{eqerror}
\end{equation}
where $\sigma_\mathrm{rms}$ is the r.m.s. noise at a velocity resolution $\delta V$
of 3.3 km/s, $\Delta V_\mathrm{CO}$ the line width.  Since the CO line was not
detected, $\Delta V_\mathrm{CO}$ is a free parameter.  Using the measurements
obtained at IRAM 30m, we obtain
\begin{equation}
\mathrm{I_{CO}} < 0.014 \sqrt{\Delta V_\mathrm{CO}}.
\end{equation}

\section{Discussion}
\begin{figure}[t]
  \begin{center}
    \FigureFile(7.5cm,9cm){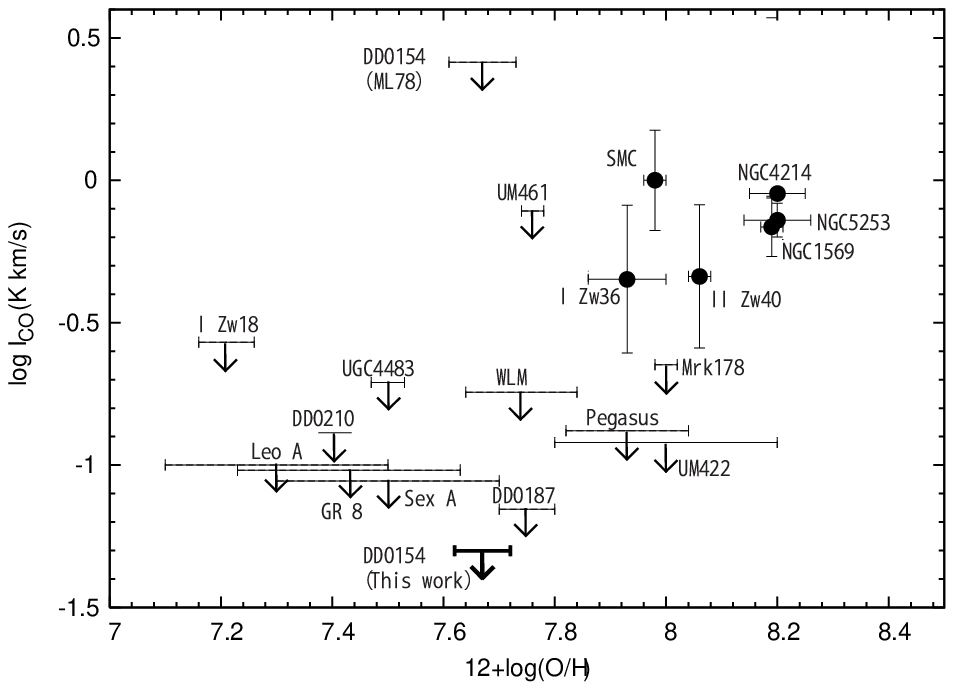}
  \end{center}
  \caption{Previous observations of CO in low-metallicity dwarfs.  Our
$3 \sigma$ upper limit on the CO intensity of DDO 154 is shown by the bold arrow.
 The metallicity and CO intensities are taken from \citet{pagel78},
 \citet{israel93}, \citet{young95}, \citet{KS96}, \citet{KS97},
 \citet{kobulnicky97}, \citet{viallefond83}, \citet{skillman93}, \citet{gallagher98},
 \citet{vanzee97}, \citet{morris78}:ML78, \citet{sage92},
 \citet{terlevich91}, \citet{skillman88}, \citet{skillman89},
 \citet{skillman94}, \citet{skillman97}, and \citet{taylor98}, following
 figure 2 in \citet{taylor98}.  For
 the previous DDO154 observation by ML78, only the $3 \sigma$ upper
 limit in antenna
 temperature $\mathrm{T_a^*}$ is given at 2.6 km/s velocity resolution;
 we converted it to intensity using an antenna efficiency of 0.5 and
 velocity width of 10 km/s.}\label{metal_Ico}
\end{figure}

A relation between molecular cloud size and CO line width exists for
Galactic molecular clouds \citep{dame86, solomon87} and
also in lower metallicity environments like the LMC \citep{mizuno01} and
NGC6822 \citep{gratier10}.  Assuming that the same relation applies to
molecular clouds in DDO154, the size of the observed HII region
($\sim 100$ pc) gives an upper limit on the line width
$\Delta V_{CO}$ of 10 km/s.  Substituting this value to equation (2)
gives $\mathrm{I_{CO}} < 0.05$.  This upper limit is the smallest yet
obtained for any low metallicity galaxy (figure \ref{metal_Ico}).  

  We use this upper limit in the
discussions that follow.  This intensity estimate should be robust, considering that
molecular gas is unlikely extended as much as the HII region itself; the
 dust seen at 24$\mathrm{\mu m}$ is unresolved, where the MIPS 24$\mathrm{\mu m}$
 resolution of $6^{\prime \prime}$ corresponds to $\sim 90$ parsec.

Using the value above and the estimated column density of 
$\Sigma_\mathrm{H_2} = 1.6\times 10^{20} \ \mathrm{cm^{-2}}$
from section 2.2, we obtain a lower limit on the
CO-to-$\mathrm{H_2}$ conversion factor of 

\begin{equation}
\frac{X_\mathrm{CO}}{[\mathrm{cm^{-2}\ (K\ km\ s^{-1})^{-1}}]} = \frac{\mathrm{N(H_2)}}{[\mathrm{cm^{-2}}]} \frac{[\mathrm{K\ km\ s^{-1}}]}{\mathrm{I_{CO}}} > 3.1 \times 10^{21}
\end{equation}

which is $\sim 10$ times the Galactic value.

Alternative to the gas column density estimation used here which is based on the
Schmidt-Kennicutt type relation (e.g., \cite{kennicutt98},\cite{komugi05}), 
we may estimate the gas column density via a
more local argument as done in \citet{verter95} and \citet{taylor01}:

\begin{equation}
\frac{\mathrm{N(H_2)}}{[\mathrm{cm^{-2}}]}=9.5\times 10^{-23}\frac{L\mathrm{(H\alpha)(erg\ s^{-1})}\tau_\mathrm{SF(yr)}}{\epsilon_\mathrm{SF}\mathrm{R(pc)^2}}
\end{equation}

where $\tau_\mathrm{SF}$, $\epsilon_\mathrm{SF}$, and R are the
timescale of star formation, star formation efficiency, and radius of
the molecular cloud, respectively.  Collisions between clouds can regulate
$\tau_\mathrm{SF}$ \citep{komugi06}, and the upper limit on the cloud
collision timescale can be estimated by the crossing time \citep{lo03},
which is often applied in local star formation arguments as done here \citep{verter95, 
taylor01}.  In DDO154, the rotation
velocity of $\sim 30\ \ \mathrm{km\ s^{-1}}$ at the optical radius of
1.4 kpc \citep{carignan89} sets $\tau_\mathrm{SF} \le 2\times 10^{8}\ \mathrm{yr}$. 
The lower limit of
$\tau_\mathrm{SF}$ is constrained by the timescale of
disruption by OB stars, which is few$\times 10^7$ yr in nearby galaxies
(e.g.,\cite{egusa04}, \cite{egusa09}).  We therefore assume
$\tau_\mathrm{SF}=10^8$ yr, in accordance with previous studies.  
For $\epsilon_\mathrm{SF}$, it has been known that the overall star
formation efficiency of clusters is few - 30\% in the solar
neighborhood, with a tendency for lower $\epsilon_\mathrm{SF}$ in
less evolved, young clusters (\cite{lada03}).  For entire molecular clouds, 
$\epsilon_\mathrm{SF}$ is observed to be lower than 10\%
in our Galaxy (e.g., \cite{duerr82}, \cite{evans91}) and also in nearby galaxies
\citep{wilsonmatthews95, taylor99}.  Even in low metallicity systems which may have higher
star formation efficiencies, $\epsilon_\mathrm{SF}=0.2$ should give a conservative upper limit estimate 
(e.g.,\cite{yasui08}).  Substituting these values and the molecular cloud radius upper limit
of $\mathrm{R}=50$ parsecs into equation (4), we obtain $\mathrm{N(H_2)} > 1.4\times 10^{20}$ 
and a corresponding lower limit of $X_\mathrm{CO} > 2.7 \times10^{21}$. 
 Although this estimate is highly uncertain considering large
 uncertainties in $\tau_\mathrm{SF}$ and $\epsilon_\mathrm{SF}$, and
 accurate to factors of few at best, the $X_\mathrm{CO}$
lower limit obtained using two different methods, the former based on
an empirical global star formation law and the latter based on local star formation
processes, are consistent at $\sim 10$ times larger than the typical
Galactic value.  The lower limit on the conversion factor is consistent with the $X_\mathrm{CO}$-metallicity relation, as shown by \cite{wilson95} and \cite{arimoto96}.  Our results, however, are inconsistent with constant conversion factor as a function of metallicity, as claimed by \cite{rosolowsky03}, \cite{bolatto04, bolatto08}, which is valid only above $12+\log[\mathrm{O/H}] \sim 8.0$.

  In order for our CO intensity upper limit to be consistent with a
Galactic conversion factor, increasing the star formation efficiency
$\epsilon_\mathrm{SF}$ to 1 is not sufficient.
The star formation timescale $\tau_\mathrm{SF}$ must also become significantly shorter, so
that the molecular cloud converts all of its gas into stars in at most
several times $10^7$ years, which seems difficult to realize.

  It should be pointed out that our observation is
done towards a single star forming region, which is unresolved by the
telescope beam.  The strong UV radiation field from the central stars in
a metal-poor environment should make the CO emitting region smaller 
\citep{maloney88}, so that
the molecular gas is distributed in a more extended way than the CO
molecules.  The Schmidt-Kennicutt law estimation of molecular gas
implicitly derives the total molecular gas surface density within the
unresolved region, including diffuse molecular gas that is not directly associated with
Region 2.   Therefore, the $X_\mathrm{CO}$ estimation given here may still be biased
towards a higher value as pointed out in \cite{wilson95}.  In order to
mitigate such biases, we have stressed the selection of conservative values whenever
possible.  The consistent lower limit on $X_\mathrm{CO}$ given from
molecular gas estimation based on a individual molecular gas argument in
equation (4), may justify the use of such methods using an empirical
star formation law.  Nevertheless, figure 4 alone should not be taken as an
indication on $X_\mathrm{CO}$ at low metallicities, but as a
representation of how CO is undetectable at such metallicities, and the
depth of the CO observation given here.

\section{Summary and Conclusions}

We have obtained sensitive spectra in search of CO in DDO154, a low
metallicity dwarf irregular galaxy.  In particular, the targeted HII
 region (Region 2) is expected to contain a detectable amount of
 molecular gas, if a Galactic conversion factor is applicable in such
 low metallicity environments.  Using two large aperture single dish
telescopes, however, we did not detect any significant CO
emission in Region 2.

  The obtained upper limit on the CO intensity is one of the
lowest in a low metallicity dwarf galaxies.  We calculate the expected
molecular column density from two views, one based on the
Schmidt-Kennicutt law extrapolated to low densities, and another based
on a local argument of a collapsing molecular cloud.  Both estimations
gave similar values.  Applying a method used previously (e.g.,
\cite{verter95}) and using conservative values in all assumptions, we
derive a lower limit on the CO-to-$\mathrm{H_2}$ conversion factor which
is at least 10 times higher than in the Milky Way.  Our results strongly support a
an increasing conversion factor with decreasing metallicity, but
inconsistent with a constant (Galactic) conversion factor, unless the 
molecular cloud forms stars very efficiently (all of
the gas converted to stars) at a timescale significantly 
shorter than $10^8$ years.  Molecular gas masses in ultra-low metallicity
systems estimated using
Galactic conversion factors may underestimate masses significantly.\\[0.5cm]

S.K. is grateful to the staffs at the NRO and IRAM, for their generous
support in the observations.  S.K. thanks N. Sakai for suggestions on 
data reduction.  We acknowledge the usage of the NASA/IPAC
Extragalactic Database (NED), which is operated by the Jet Propulsion
Laboratory, Caltech, under contract with the National Aeronautics and
Space Administration.  S.K., C.Y., and B.H. were supported by the
Research Fellowship from the Japan Society for the Promotion of Science
for Young Scientists.

\end{document}